\documentclass[twocolumn,showpacs,preprintnumbers,amsmath,amssymb]{revtex4}

\usepackage{graphicx}
\usepackage{dcolumn}
\usepackage{bm}

\begin{document}


\title{Chaos-induced breaking of the Franck-Condon approximation}
\author{Hisatsugu  Yamasaki}
\email{hisa@physics.s.chiba-u.ac.jp}
\homepage{http://zeong.s.chiba-u.ac.jp/~hisa/}
\author{Yuhei Natsume}%
\affiliation{%
Graduate School of Science and Technology, Chiba-University\\
Inage-ku, Chiba, 263-8522 Japan
}%

\author{Akira Terai}
\author{Katsuhiro Nakamura}
\affiliation{
Department of Applied Physics,
Osaka City University\\
Sumiyoshi-ku
Osaka, 558-8585 Japan
}%
     
\date{\today}

\begin{abstract}
We investigate the vibrational structure of electronic spectra for the
transition from the non-degenerate $A$
state to $E$ states in $E_g\otimes e_g$ Jahn-Teller systems with the
trigonal field included. In connection with the underlying chaotic behavior
for vibronic energy levels in this model reported by the present authors in
Phys.Rev.{\bf E68}(2003)046201, we extend the analysis by Longuet-Higgins et
al. to the classically chaotic system. In particular, the
triple-humped structure is manifest with increasing the anharmonicity. Such
 structure is completely inconsistent with the shape obtained from
 Franck-Condon(FC) approximation, and is caused by the chaos-induced beaking
 of the FC principle. 

\end{abstract}

\pacs{05.45.Mt,71.70.Ej,82.90.+j,03.65.-w,31.30.Gs}
\maketitle

%
%
We investigate the vibronic problem in degenerate $E_g$ orbitals of
$d$-levels in transition metal ions coupled with 2-d vibrational modes
$e_g$ expressed by coordinates $Q_1(=Q_{e_gu})$ and $Q_2(=Q_{e_gv})$ by taking into account the
trigonal distortion. This $E_g\otimes e_g$ model
is the typical system\cite{4} showing dynamic Jahn-Teller effects(DJTE),
which has been discussed in the field of magnetism for transition-metal
ions. In fact, the numerical work was launched \cite{7,8} by
Longuet-Higgins et al[LH] in 1958. Since then, the double-humped vibrational structure of
electronically allowed transitions from the electronically non-degenerate
ground state $A$ to a Jahn-Teller degenerate state $E_g\otimes e_g$ has been the
fundamental subject in optical properties of transition-metal ionic
compounds. There the ground state is taken as the level of zero-phonon in $A$ state
at zero temperature.

%
%
Recently, the present authors\cite{32} found the relationship between
the chaotic behavior of this system and the magnetic $g$-factors of
electronic orbital angular momentum as well as
features of level statistics for vibronic states. The
statistical properties of those levels and the energy dependence of
$g$-factor shed light on the quantum signature of ``Chaos'' in this
Jahn-Teller system.

%
%
In the present paper, we first report the novel vibrational structure in the spectra of
the transition from the non-degenerate $A$ state to the excited states in the
$E_g\otimes e_g$ model. Secondly, the comparative study between vibrational
structure in quantum mechanical treatment and the semiclassical one in Condon approximation for the
adiabatic potential is made in order to see the role of ``Quantum
Chaos''\cite{1,2}. 

%
%
The Hamiltonian matrix $\mathcal{H} $ is expressed as 
\begin{eqnarray}
 \mathcal{H} &=& -\frac{\hbar^2}{2}\left(\frac{\partial^2}{\partial Q^2_1}+\frac{\partial^2}{\partial Q^2_2}\right){\bf{I}} 
 + \frac{1}{2}\omega^2(Q_1^2+Q_2^2){\bf I} \nonumber \\
 &+& k[-Q_1\sigma_3+Q_2\sigma_1]+bQ_1(Q_1^2-3Q_2^2){\bf I},\label{eqn:1}
\end{eqnarray}
where $\bf{I}$ is the unit matrix and $\sigma_i$ with $i=1,2,3$ are Pauli matrices. 
This Hamiltonian operates on $\varphi_u$ and $\varphi_v$ bases in $E_g$
state, which are expressed as $|3z^2-r^2\rangle$ and
$|x^2-y^2\rangle$, respectively. In the third term called ``linear
Jahn-Teller matrix'', $k$ is the coupling between electronic and vibrational
states. The strength of the anharmonic trigonal field is expressed by $b$ of the forth term.

%
%
Without the trigonal field, the second and the linear Jahn-Teller matrix give
the adiabatic potential energy surfaces(APES) for axial symmetry: APES is
expressed as $(1/2)\omega^2\rho^2 \pm k\rho$, where
$\rho=\sqrt{Q_1^2+Q_2^2}$. It should be noted that APES is independent on 
azimuthal angle
$\theta=\tan^{-1}(Q_2/Q_1)$. Therefore, APES for the lower branch has
the continuous minima whose value is $k^2/\omega^2$. Such minima draw the
circle located at the bottom of the so-called ``Mexican Hat''
potential. 
%
%
In the vibronic problem, we take into account the first term of
(\ref{eqn:1}) in addition to the other static terms. Namely, we treat the vibrational
modes $e_g$ as the 2-dimensional harmonic oscillator for quantum mechanics in
order to investigate the dynamical effects: Wavefunctions are described as
$\phi_{nm} = F_{n|m|}(\rho)e^{im\theta}$, where $n=1,2,\ldots$ and
$m=n-1,n-3,\ldots,n+1$. Here, $F_{n|m|}(\rho)$ is the confluent
hypergeometric function\cite{11,25}. According to the conventional procedure
for the degenerate electronic state $E_g$, we employ the expressions
$\varphi_{\pm}$ of eigenfunctions for 2-dimensional angular momentum, 
which are transformed from $\varphi_u$ and $\varphi_v$: The
transformation is $\varphi_{\pm}=(1/\sqrt{2})(\varphi_u\pm
i\varphi_v)$. Thus, the linear Jahn-Teller matrix in (\ref{eqn:1}) is
transformed to $\sqrt{2}k\rho e^{i\theta} i\sigma_2$. Using
$\phi_{nm}(\rho,\theta)$ and $\varphi_{\pm}$, we can express bases of present
vibronic wavefunctions as 
\begin{equation}
 \Phi_{nm}^{\pm}=\varphi_{\pm}\cdot \phi_{nm}(\rho,\theta). \label{eqn:2}
\end{equation}
The qunatum mechanical expression of the present $E_g\otimes e_g$ model 
is obtained from the
following nonvanishing elements:
\begin{align}
& \langle\phi_{n,m}|\rho
 e^{i\theta}|\phi_{n',m'}\rangle \nonumber \\
&=
 \{\frac{\hbar}{2\omega}\left[n\pm(m-1)\right]\}^{1/2}\delta_{n',n\mp1}\delta_{m',m-1} \nonumber \\
& \langle\phi_{n,m}|\rho
 e^{-i\theta}|\phi_{n',m'}\rangle \nonumber \\
&=
 \{\frac{\hbar}{2\omega}\left[n\pm(m+1)\right]\}^{1/2}\delta_{n',n\pm1}\delta_{m',m+1}. \label{eqn:21-2}
\end{align}
If we assign the quantum numbers $j=\pm 1$ to $\Phi_{n,m}^{\pm}$, the
Jahn-Teller interaction
without the anharmonic term connects the states with the same quantum number,
$\ell=m-(1/2)j\quad(j=\pm 1)$. The present matrix decomposes
into matrices labeled by quantum number $\ell$. For any given value of
$\ell$, $m$ can take two values, $m=\ell-1/2$ and $\ell+1/2$ corresponding
to $j=-1$ and $+1$, respectively. Thus, the $p$-th eigenfunction for a given
$\ell$ is expressed as $\Psi_{p,\ell}$.
As a results, the total angular momentum $\ell$ whose values are
$\pm1/2,\pm3/2,\pm5/2,\ldots$ becomes the good quantum-number in the case of $b=0$.
\begin{figure}[h]
  \begin{center}
   \resizebox{0.42\textwidth}{!}{\includegraphics{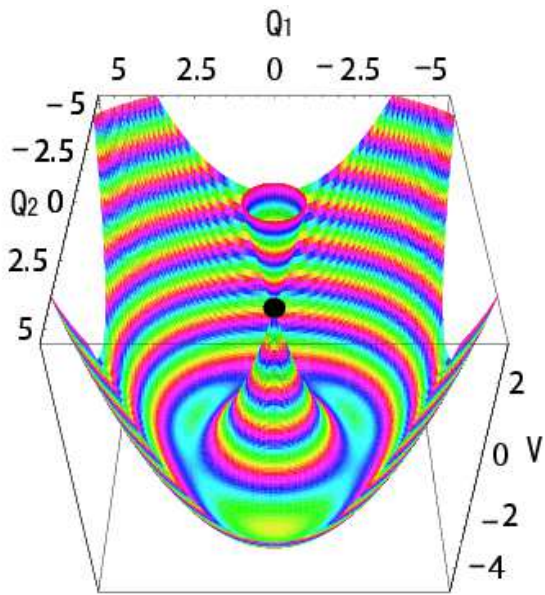}}
  \end{center}
  \vspace*{-7mm}
  \caption{APES with the trigonal distortion expressed as
 $b Q_1(Q_1^2-3Q_2^2)$. Three minima are induced. The full point means
 the conical intersection.}
 \label{fig2}
\end{figure}

%
%
In the absence of the trigonal field in (\ref{eqn:1}),
LH obtained the vibrational structure appearing in allowed electronic
spectra from the non-degenerates electronic state $A$ to $E_g\otimes e_g$
system for various values of $k$, finding the
structure with two intensity maxima in the energy region of $-10\hbar\omega
\leq \varepsilon \leq 15\hbar \omega$. This double-humped structure can be explained by the
transitions to a pair of branches of APES in the $E_g\otimes e_g$ system.

\begin{figure*}[htbp]
  \begin{minipage}{.47\textwidth}
   \includegraphics[width=\linewidth]{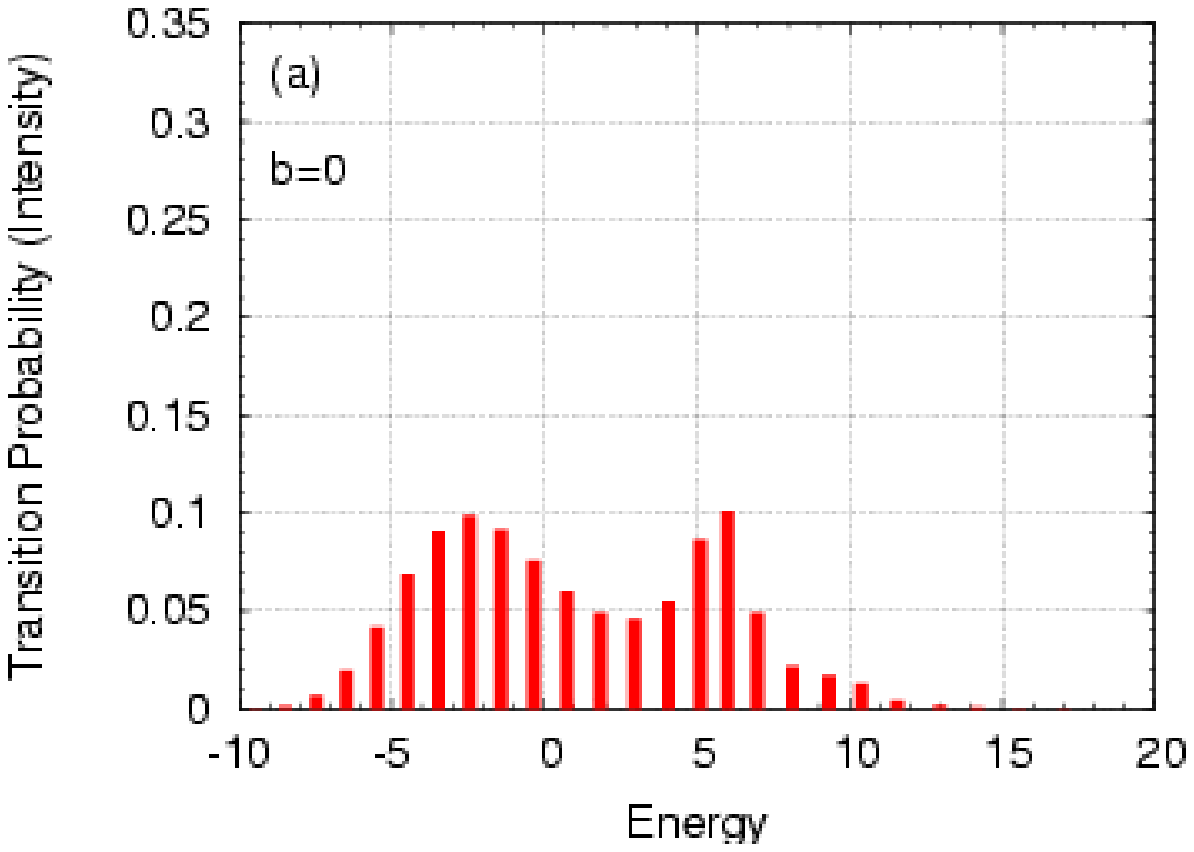}
   \includegraphics[width=\linewidth]{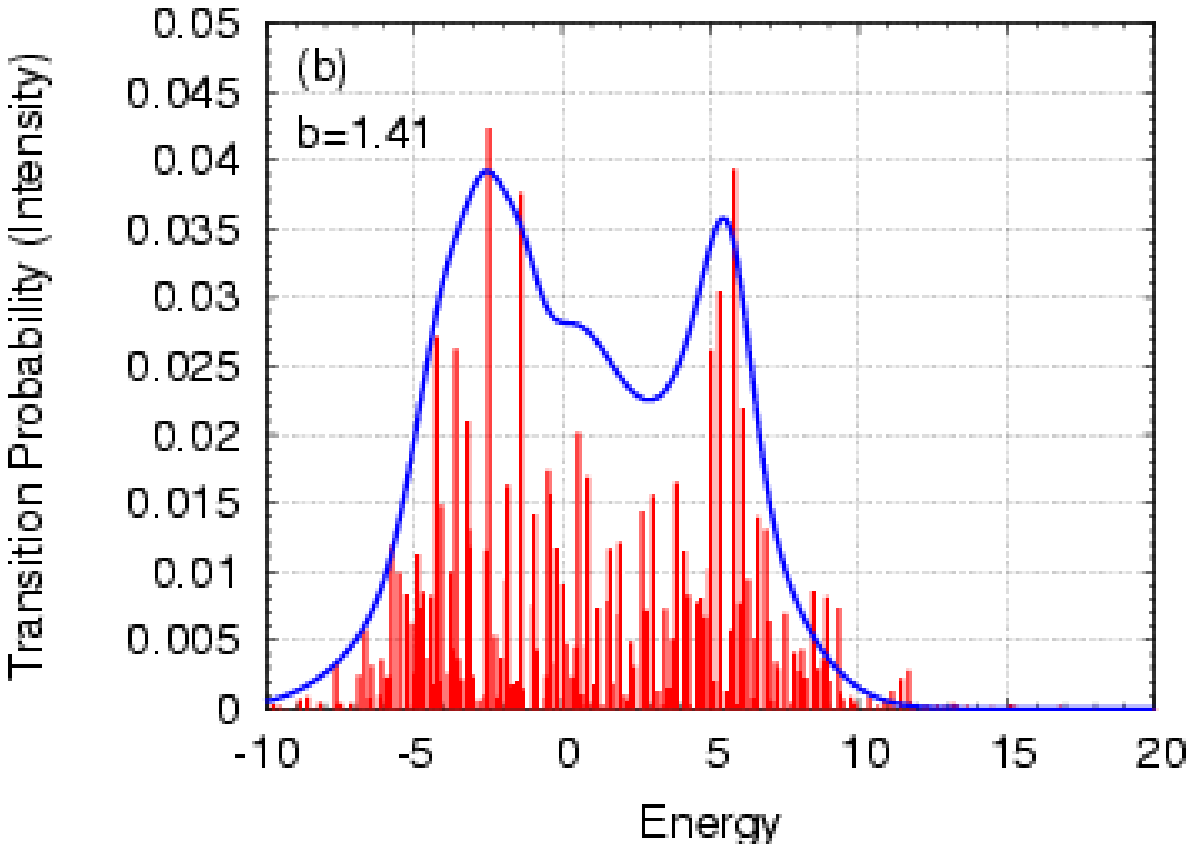}
  \end{minipage}
\hfill
  \begin{minipage}{.47\textwidth}
   \includegraphics[width=\linewidth]{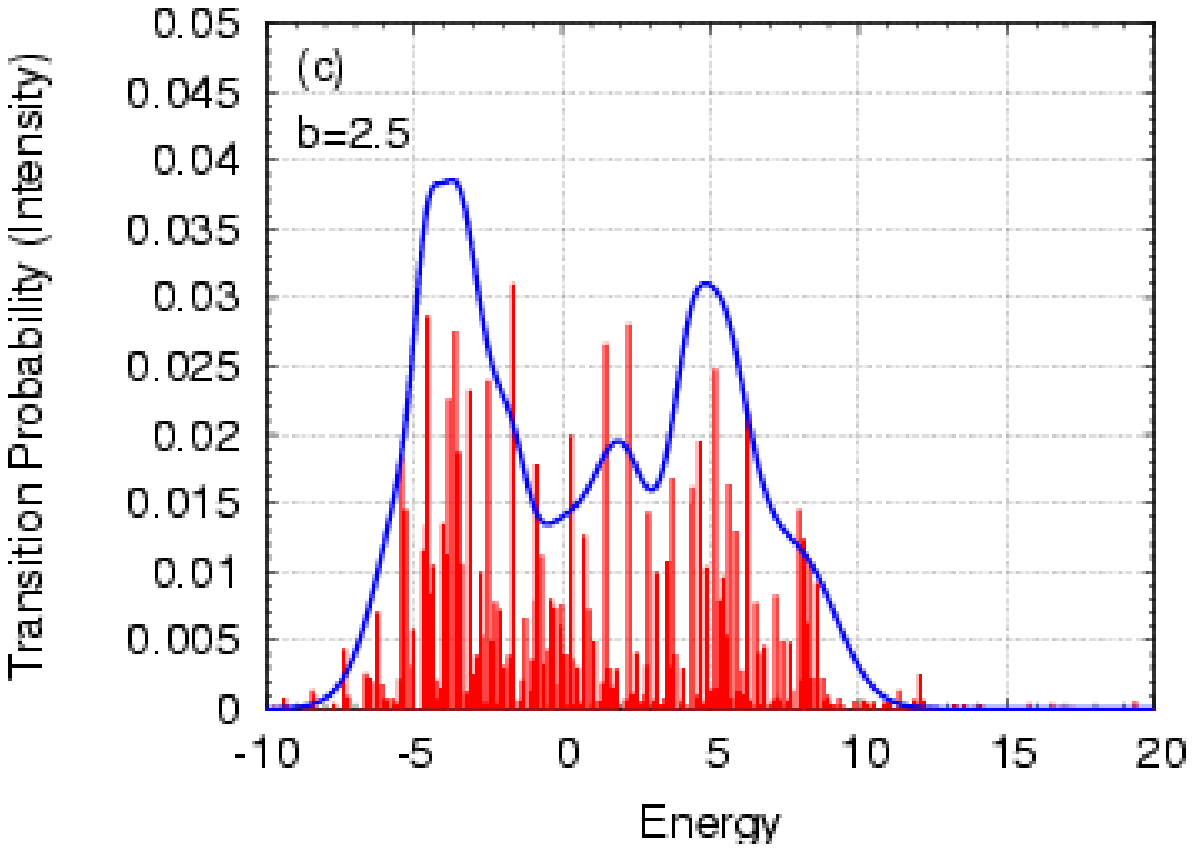}	
   \includegraphics[width=\linewidth]{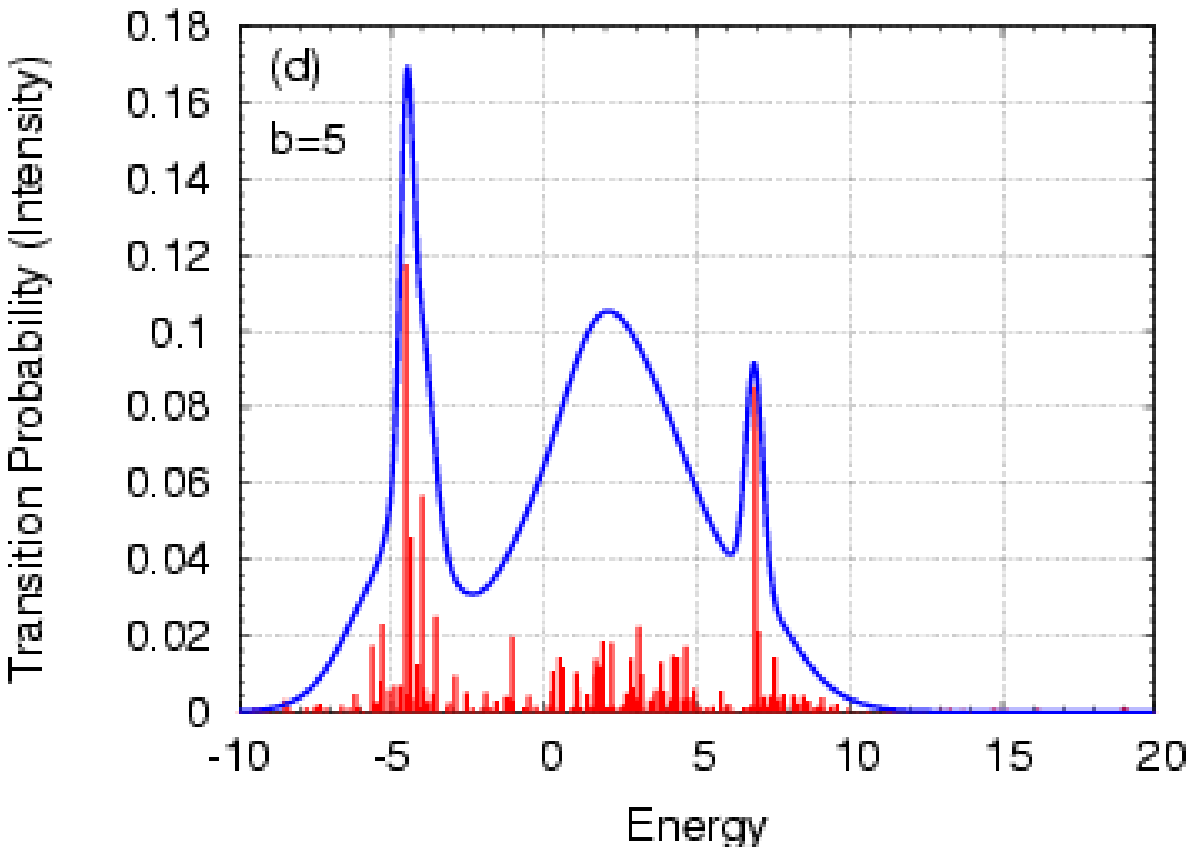}
  \end{minipage}
  \caption{The calculated vibrational structure in transition spectra from $A$ to
$E_g\otimes e_g$ system with the trigonal field are shown in
(a)(b)(c)(d), where the strength of the field $b$ is 0,1.41,2.5
and 5, respectively. The coupling $k$ between electronic and vibrational states
 is fixed to be 4.47. The envelop functions for line-spectra are drawn by
 solid lines. } 
\label{fig3}
\end{figure*}
In this paper, we consider the third term $(b \neq 0)$ in (\ref{eqn:1}), which expresses
the anharmonicity keeping invariant to any operation in the cubic group. In fact,
it gives the trigonal ligand field $\rho^3\cos{3\theta}=Q_1(Q_1^2-3Q_2^2)$ to the adiabatic potential as shown in
Fig.\ref{fig2}: The continuous circular symmetry is destroyed and three minima
appear in the lower branch. We shall calculate
vibrational structures in the spectra in this case.
O'Brien investigated\cite{3} the system (\ref{eqn:1}) with the trigonal field in the
low-energy approximation that $\rho$ is fixed to be $\rho_0$ at the bottom of
the Mexican-hat, but gave no discussion on the spectra\cite{3}. Here we numerically calculate eigenvalues and
eigenvectors without having recourse to such an approximation in order to
obtain the spectra exactly. In this case,
the angular momentum $\ell$ is not a good quantum number: By the trigonal
field, the levels for $\ell=1/2\pm 3\nu$ (where
$\nu=1,2,3,\ldots$) are mixed into the levels of $\ell=1/2$. We
get in this way a set of vibronic doublets arising from combinations with
$\ell=\pm1/2,\mp5/2,\pm7/2,\mp11/2,\pm13/2,\ldots$ (where either the upper or
lower sign is to be taken throughout). Namely, we treat the doubly degenerate
$E$ representations of the symmetry group $C_{3v}$. On the other hand,
solutions with $\ell=\pm 3/2,\pm 9/2,\pm 15/2,\ldots$ correspond to identical
representation $A$ and $B$ of group $C_{3v}$: $A$ is the mode with the symmetric
composition for $\ell > 0$ and $\ell < 0$, while $B$ with anti-symmetric one.

One of the observable phenomena where the trigonal field plays an
notable role is the spectral line
shape of a parity-allowed transition in which the final state is $E_g\otimes
e_g$ while the initial state is an orbital singlet $A$.
The quantum mechanical calculation of the spectral line-shape is possible if the
eigenvectors of the matrix with elements
(\ref{eqn:21-2}) are known. In short, the spectra are made of the probability
density to find the first basis $\Phi^{-}\equiv u_{-}({\bf
r})\phi_{1,0}(\rho,\theta)$ in the final state of the transition.

%
%
The calculated vibrational structure in the transition spectra from $A$ to
the excited levels are shown in
Fig.\ref{fig3}(a),(b),(c) and (d), where the strength of the trigonal field $b$ is 0,1.41,2.5
and 5, respectively. 
Here, the coupling parameter $k$ in (\ref{eqn:1}) is fixed
to be 4.47 (i.e., $k^2=20$). It is certain that the double-humped structure
for $b=0$(see Fig.\ref{fig3}(a)),
composed of transitions only to levels in $\ell =1/2$,
agrees with that in LH\cite{7}. 
\begin{figure}[htbp]
  \begin{center}
   \resizebox{0.42\textwidth}{!}{\includegraphics{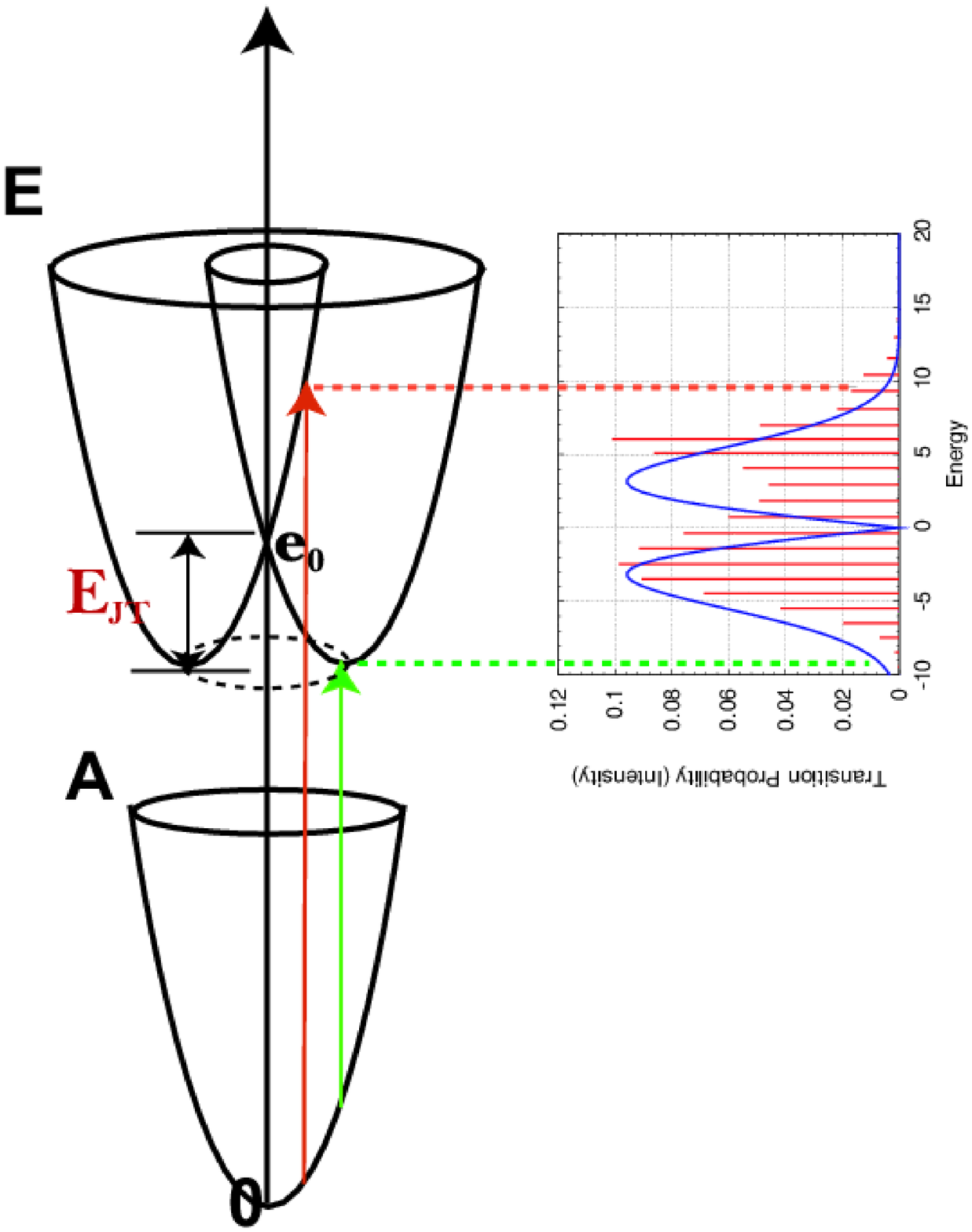}}
  \end{center}
  \vspace*{-7mm}
  \caption{APES of the electronic states $A$ and
 $E$. The spectral line-shape function for the transition from $A$ to $E$
 states obtained by the semiclassical Franck-Condon approximation is shown by
 the solid curve for the typical case of $b=0$, which has been reported in
\cite{31}. This function becomes zero at $\varepsilon = 0$ from mathematical
 nature of APES at $\rho=0$. The corresponding vibronic structures are also drawn by
 dotted lines.}
 \label{fig7}
\end{figure}
As $b$ increases, however, the double-humped structure changes into the triple-humped one, 
because of the mixing of levels for
$\ell=\mp5/2,\pm7/2,\mp11/2,\ldots$ into the excited $E$ state. In other
words, an extra
new hump appears at the conical interaction, i.e., around the dip of
double-humped structures. The central branch in the spectrum is the
common result proper to the system with the trigonal field, as shown in
Fig.\ref{fig3}(b),(c) and (d). This new hump can also be
interpreted as ``Quantum Chaos'' emanating   
in the region $b\lesssim k$\cite{32}, as explained below.

%
%
\begin{figure}[htbp]
  \begin{minipage}{.49\textwidth}
   \includegraphics[width=\linewidth]{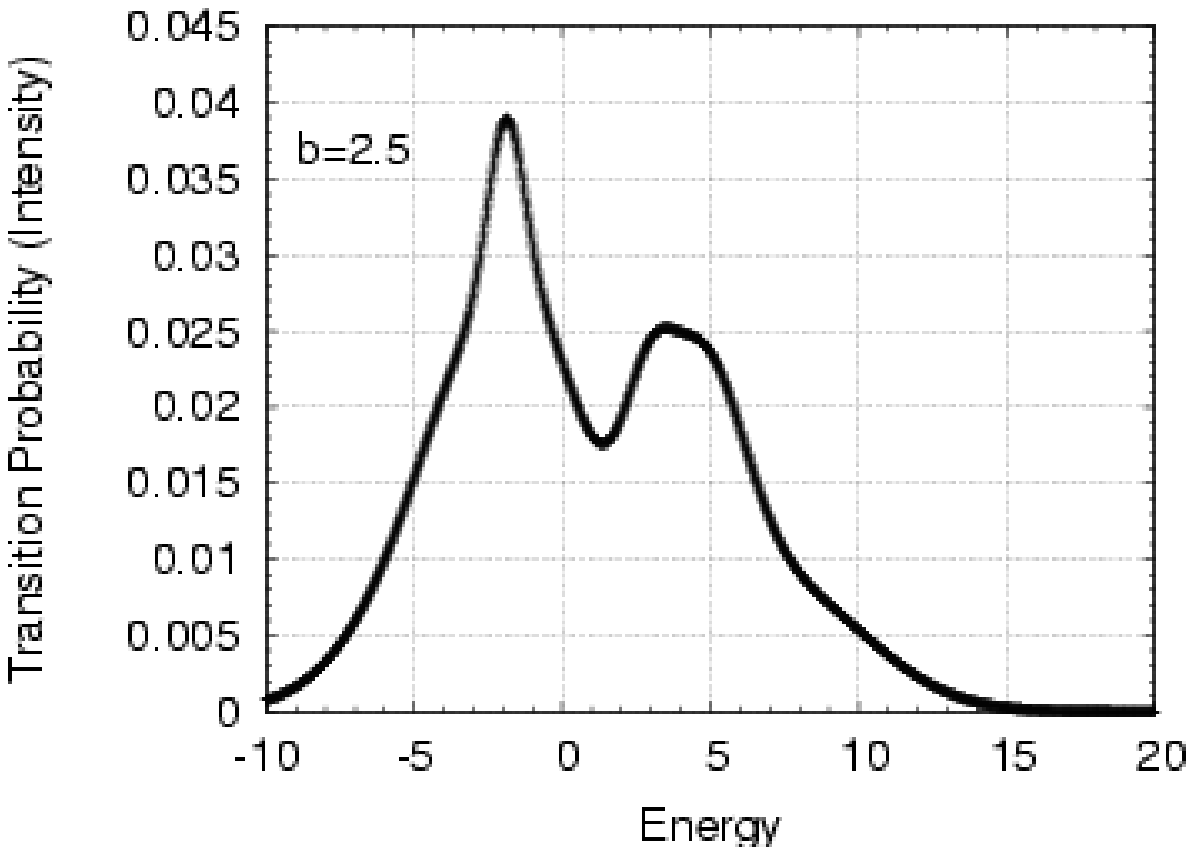}
  \end{minipage}
\caption{The shape of spectra for $k=4.47$ and $b=2.5$ obtained by
 Franck-Condon approximation discussed in the text. Temperature is determined
 to be $0.72\hbar\omega/\kappa$. The shape is completely inconsistent with
 feature in Fig.\ref{fig3}(c).}
\label{fig4}
\end{figure}
In order to discuss this triple-humped structure in detail, we calculate
semiclassical spectra in Franck-Condon(FC) approximation, which is schematically
shown in Fig.\ref{fig7}. The physical implication for this approximation is
intuitively clear: It shows that the electronic transition takes place so
rapidly that the nuclear positions do not change during the transition. 
In fact, the two intensity maxima of the transition spectra without
anharmonicity were revealed  by
Y.Toyozawa and M.Inoue\cite{31}. They showed the absorption line-shape function for
the transition spectrum by using the semiclassical FC
approximation. Here, we shall show the consequence of the FC
approximation applied explicity to the system with cases of $b\neq
0$. Namely, APES consists of two branch surfaces as shown
in Fig.\ref{fig2}, corresponding to 
\begin{equation}
 \varepsilon_{\pm}(\rho,\theta)=\varepsilon_0 \mp k\rho+b\rho^3\cos{3\theta}\label{30}
\end{equation}
where $\varepsilon_0$ is the excitation energy from the ground state $A$ to
the excited one $E$ at $\rho=0$.
Within the semiclassical FC approximation, the normalized line-shape
function of the optical absorption is given by
\begin{eqnarray}
 F(\hbar\omega)&=&\frac{1}{2}\sum_{\pm}\int\int d\theta \sin{\theta} d\rho \rho\left(\frac{1}{\pi \kappa T}\right)\nonumber \\
&&\times\exp\left(-\frac{\rho^2}{\kappa T}\right)\delta\{\hbar\omega-\varepsilon_{\pm}(\rho,\theta)\},
\end{eqnarray}
where $\kappa,T$ and $\hbar\omega$ are the Boltzmann constant, the absolute
temperature and the photon energy, respectively. Namely $\kappa T$ give the
width to line spectra. In the present calculation, we use $\kappa
T=0.72\hbar \omega$. Without anharmonicity the
absorption shape function is integrable\cite{31}.($b=0$):
This gives a line-shape which is completely split into upper and lower
parts, as shown in Fig.\ref{fig7}. In Fig.\ref{fig4}, we show the line
shape in FC approximation for $b=2.5$, which is completely
inconsistant with the triple-humped feature in Fig.\ref{fig3}(c).

In the presence of the
trigonal(anharmonic) potential, the nonadiabatic motion of electron wave
packet can show chaotic
features with the outstanding occupation probability at the conical
intersection due to its subtle distortions. This means the increase of density of state at the energy
where the conical intersection locates. Therefore, the transition from $A$ state has
extra final states at the level of the conical intersection, i.e.,
triple-humped structure as a whole.  Here a simple
application of the FC principle is not justified(see the
pertinacious double-humped structure in Fig.\ref{fig4}). 

%
%
As a result of comparative discussion between vibrational structure and the 
semiclassical spectra in FC approximation, we would
like to point out the following fact:
The new peak comes from the non-adiabatic dynamic mixing to the
pair of lower and upper APES for electronic $E_g$ state, which is a typical
quantum manifestation of the underlying chaos.

In conclusion, we find
a new hump induced by the trigonal field in the transition spectrum, which is manifesting of quantum chaos
in vibrations coupled with electronic states in the case of $b=0$. In the strong coupling limit
($k\gg 1$), the void between a pair of APES is obvious. This fact led to the double-humped structure of
LH\cite{7} in the nonadiabatic spectra of the transition from $A$ to
degenerate $E_g\otimes e_g$ states. This spectra can be nicely reproduced by application
of the FC principle. However, the anharmonic trigonal term, which
yields the new density of states near the energy of the conical intersection,
breaks the FC principle. 

In the case of
the oscillation of magnetic $g$-factor\cite{32} we assumed the
small $k$ value. The quenching of the regular oscillation was also caused by the underlying classical chaos due to anharmonic term. On the
contrary, in the present work, we require the $k$ value to be sufficiently
large to ensure the double-humped transition spectra due to the obvious splitting of
two APES, and chaos in the nonadiabatic electron wave packet plays an essential
role. 

The relationship between vibrational structures of spectra and chaotic
behavior in the present quantum system is quite attractive, though there is 
very much left to study of analyzing this characteristic structure
in connection with experimental work for transition-metal ions in compounds.

\end{document}